\documentclass[amsmath,showkeys,superscriptaddress,onecolumn,12pt]{revtex4-2}
\usepackage{float}
\usepackage{rotating}
\usepackage{tabularx}
\usepackage[table]{xcolor}
\usepackage{float}
\usepackage{multirow}
\usepackage{hhline}
\usepackage{colortbl}
\usepackage{titlesec}
\usepackage{amsmath}
\usepackage{amsmath}
\usepackage{amssymb}
\usepackage{graphicx}
\usepackage[utf8]{inputenc}
\usepackage[colorlinks=true,linkcolor=blue,citecolor=blue,urlcolor=blue,breaklinks]{hyperref}
\begin{document}
\title{Demonstration of entanglement and coherence in GHZ-like state when exposed to classical environments with power-law noise}
\author{Atta Ur Rahman}
\email{Attapk@outlook.com}
\author{ZhaoXu Ji}
\email{jizhaoxu@whu.edu.cn}
\author{HuanGuo Zhang}
\email{liss@whu.edu.cn}
\address{Key Laboratory of Aerospace Information Security and Trusted Computing,
Ministry of Education, School of Cyber Science and Engineering, Wuhan University, China}
\begin{abstract}
Entanglement and coherence protection are investigated using the dynamical map of three non-interacting qubits that are initially prepared as maximally entangled GHZ-like states coupled to external fields in solid-state and superconducting materials. Thermal fluctuations and resistance in these materials produce power-law (PL) noise, which we assume controls external fields in three different configurations with single or multiple noise sources. The genuine response of isolated environments to entanglement and coherence retention is analyzed. We also briefly discuss the initial purity and relative efficiency of the GHZ-like states. Unlike the multiple PL noise sources, exposure, the GHZ-class state remains partially entangled and coherent for an indefinite time when subject to single noise source. However, long-term non-local correlation and coherence are still feasible under multiple noise sources. Due to the lack of back-action of the environments, the conversion of the free state into the resource GHZ-class states is not allowed. The parameter optimization, in addition to the noise phase, regulates disorders, noise effects, and memory properties in the current environments. Unlike the bipartite states and tripartite W-type states, the GHZ-like state has been shown to have positive traits of preserving entanglement and coherence against PL noise. 
\end{abstract}
\maketitle
\section{Introduction}
Realistic quantum systems have played an increasingly important role in defining and promoting quantum technologies in recent years, and they have been extensively investigated \cite{it, tech}. They are both real and non-isolated in the sense that their correlations are constantly interacting with huge quantum systems in their surroundings, which are referred to as environments. These environments have a prominent place in quantum research disciplines, and they are studied both theoretically and experimentally \cite{theo,exp1,exp2,exp3}. To pursue this purpose, investigation of the quantum correlation and coherence of a quantum system is essential for quantum information processing. The characterization of non-local correlation present among the constituents of a system is done by investigating the quantum entanglement \cite{ent1,ent2,ent3}. The concept of entanglement was first introduced by Einstein \cite{einstn1,einstn2} and then Schr\"{o}dinger \cite{schro} which increased the performance of quantum information processing and other related protocols \cite{qct1,qct2}. The realisation of violation of the Bell inequalities for a quantum system led to the conclusion that the entanglement of a system is a pure non-local phenomenon. This established that no classical means can produce the entanglement \cite{nonlocal}. This non-local phenomenon gives a quantum system its distinguishing feature of being able to be described only as a whole, rather than in its parts. Non-local correlations are distinguished from typical local correlations by entanglement \cite{ccr,ccr1}, which is an essential resource. Besides, coherence is a necessary condition for a quantum system to be entangled, and a decoherent state is sometimes referred to as disentangled. Entanglement and coherence are thus equally crucial for quantum information sciences to be utilised in practice.\\
We focus on entanglement and coherence, which are useful to determine the quantum correlation of a non-local system. Ideally, one can hope that an environment must remain in a non-Markovian regime for a sufficiently long time to permit the desired quantum information tasks to be deployed. Such quantum information processing tasks are minimized by the external noisy effects of an environment which unexplores the information to be shared \cite{int1,int2,int3,int4}. These detrimental effects for the survival of the entanglement have great significance in the dynamics of open quantum systems \cite{fatal1,fatal2,fatal3}. This interaction between the quantum state and environment in some cases when increases, entanglement and coherence decreases. The interaction picture of a quantum state with its environment is of two types, the classical and quantum interaction picture \cite{clasical, quantum}. The classical interaction picture has more significance than the quantum one. The reason is that it is more favourable and accurate for the system-environment coupling and has many degrees of freedom. The entanglement fading effects are investigated widely through theoretical \cite{Yu, Benedetti-color noise, Javed, th4} and experimental \cite{expe} means for the dynamics of the quantum correlation and coherence. Such effects are studied for different quantum states such as Bell's \cite{bell1,bell2}, GHZ-like \cite{GHZ1,GHZ2}, and W-like states \cite{w1}. These states are further studied in subjection to different kinds of environments generating distinct noises like the random telegraph, static, Ornstein Uhlenbeck, and mixed classical noises \cite{Javed, Kenfack}.\\
In this paper, we investigate the dynamics of entanglement and coherence for three non-interacting qubits, initially prepared as maximally entangled GHZ-like state. Our physical model is mainly focused on the Gaussian stochastic process and is defined by the power-law auto-correlation function \cite{gau1,gau2}. This results in a Gaussian kind of noise known as PL noise \cite{pw1,pw2,pw3}. The dynamics of a quantum system under such conditions can cause superposition with the noise phase and may cause decoherence and separability. In this framework, the characterization and estimation of the noisy parameters of the auto-correlation function of PL noise and the related influence on the evolution of the GHZ-like state will be investigated. Such estimations can describe the qualitative dynamics and provide solutions to avoid entanglement and coherence losses. Besides that, we also investigate the detrimental effects of a varying number of noisy sources and environments. For example, we considered three different configurations, namely common, bipartite and tripartite classical environments. In the first case, the system is exposed to one common environment with a single noise source. In the second place, the three non-interacting qubits are coupled with two classical environments with two noise sources. In the final case, each qubit is coupled with an independent classical environment each with an independent PL noisy source. This will demonstrate the characterization of environments and the number of noisy sources to favour entanglement and coherence protection. The noise parameter optimization will also be extensively studied in this work. In this context, we aim to provide proper fixing and tuning of the PL noise parameters for the survival of longer non-local correlation, coherence and extended preserved memory properties. In this regard we will be using quantum negativity and entanglement witness to evaluate the entanglement dynamics in the GHZ-class state. For the coherence and information decay , we will deploy purity and von Neumann entropy approach.\\
This paper is organized as: In Sec.\ref{Model and dynamics}, we give the details of our physical model, the corresponding dynamics along entanglement and coherence measures. Sec.\ref{Results and discussion} explore the analytical results obtained for the physical model. In Sec.\ref{Conclusion}, the paper closes with main concluding remarks.
\section{Model and dynamics}\label{Model and dynamics}
Our physical model consists of three non-interacting qubits that are first created as maximally entangled GHZ-like states that are distinguished by the population of excited states. Three different configurations resulting from the possible combination of the three qubits, environments, and noisy sources are also studied. The first is the common (COM) configuration, which links all three qubits to a single environment and PL noise source. In the second case, the qubits are exposed to two classical environments with two independent noise sources and is named as bipartite (BIP) configuration. In the final case, each qubit is exposed to an independent source of PL noise and environment, which is referred to as tripartite (TRI) configuration.
\begin{figure}[ht]
\includegraphics[width=16.6cm, height=7cm]{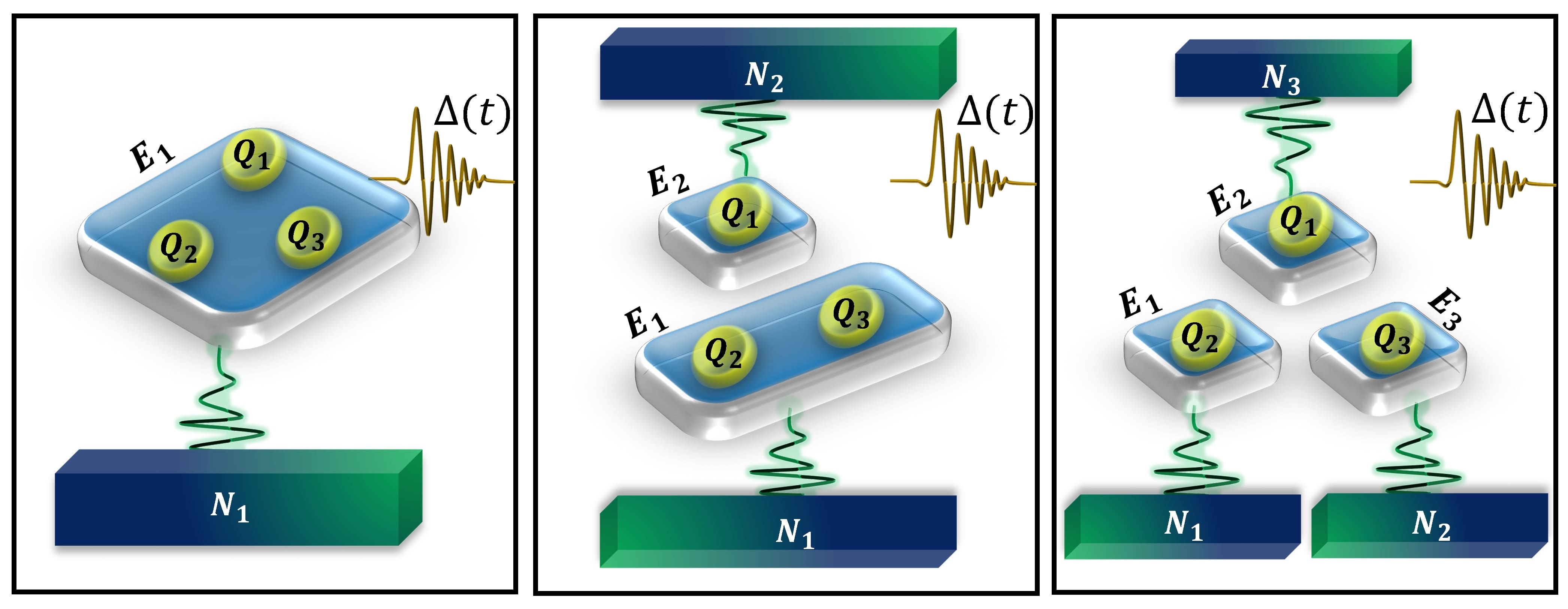}
\caption{The schematic diagram for the configuration models utilized in this study for three non-interacting qubits $Q_1$, $Q_2$ and $Q_3$ coupled to classical environments, namely, $\hbox{E1}$, $\hbox{E2}$ and $\hbox{E3}$. The classical environments are driven by the power noisy sources, $N_1$, $N_2$ and $N_3$. The similar symmetry of the qubits reveal that the three non-interacting qubits are prepared in a single GHZ-class state. The lines from the boxes demonstrate the effect of PL noise on the entanglement and coherence exhibited by the qubits. The brown lines indicates the time evolution of the three qubits with reduced amplitude represents the corresponding entanglement and coherence loss along with $\Delta(t)$, the stochastic parameter of the external fields flipping between $\pm1$ and $\Delta(t) \in \{\Delta_1(t), \Delta_2(t), \Delta_3(t) \}$. Under the effects of power-law noise, the three non-interacting qubits are coupled to common (left), bipartite (middle), and tripartite configuration (right).}
\end{figure}
This stochastic Hamiltonian model, which governed the current physical model, is written as \cite{Lionel}:
\begin{equation}
H(t)=H_1(t) \otimes I_2 \otimes I_3+I_1 \otimes H_2(t) \otimes I_3+I_1 \otimes I_2 \otimes H_3(t),
\end{equation}
where $ H_i(t) $ is the Hamiltonian of the individual sub-system and is defined as $H_i(t)=\delta I_i+\lambda \Delta(t)\sigma_i^x$ with $ i \in \{1,2,3\}$. Here, $\delta$ represents the equal energy splitting and $\lambda $ is the system-environment coupling constant. $ \Delta_i(t) $ is the stochastic parameter which introduces the classical noise to the phase of the system while $ \sigma_i^x $ and $ I_i $ are the Pauli and identity matrices respectively acting on the subspace of the system. For the time evolution of the tripartie states, we use the time unitary operator, defined as, $U(t)=\exp[-i\int^t_{to}{H(x)}dx]$ with $\hbar=1$ \cite{Lionel} . The time evolved density matrix $\rho(\Delta,t)$ for the three non-interacting qubits is given by \cite{Lionel}:
\begin{equation}
\rho(\Delta,t)=U(\Delta,t)\rho_o U(\Delta,t)^{\dagger},\label{final density matrix}
\end{equation}
where $ \Delta \in \{ \Delta_1,\Delta_2,\Delta_3 \} $ and $ \rho_o $ is the initial density matrix of the GHZ-like state and is given by:
\begin{equation}
 \rho_o=p \vert GHZ\rangle\langle GHZ\vert+\frac{I_8(1-p)}{8},\label{initial denisty matrix state}
\end{equation} 
where, p represent purity factor of the GHZ-class state and the maximum pureness of the system occurs at $p=1$. Here, $\vert GHZ \rangle=\frac{1}{\sqrt{2}}(\vert 000 \rangle+\vert 111\rangle)$.
\subsection{Power-law noise}
This section evaluates application of the power-law (PL) noise \cite{Benedetti}. The phase of the system can be represented as $ \varphi_i(t)=n \lambda \Delta_i(t)$. With classical noise, for the stochastic process to be included, $\beta$-function is to be defined, which reads as \cite{Benedetti}:
\begin{equation}
\beta(t)=\int_0^t \int_0^t K(s-s^{\prime})ds ds^\prime. \label{Beta function}
\end{equation}
The noise phase is linked to the classical field and the system through the auto-correlation function and is given by \cite{Benedetti}:
\begin{equation}
K(t-t^{\prime},\omega,\ell,\alpha)=\frac{\ell \omega (\alpha-1)}{2[1+\ell \vert t-t^{\prime}\vert]^2},\label{AC of PL}
\end{equation}
Here, we assume the dimensionless quantities $g=\frac{\ell}{\omega}$ with $\tau=\omega t$. By inserting the auto-correlation function from Eq.\eqref{AC of PL} in Eq.\eqref{Beta function}, one gets the $\beta$-function as \cite{Benedetti}:
\begin{equation}
\beta(\tau)=\frac{g\tau(\alpha -2)-1+(1+g \tau)^{2-\alpha }}{(\alpha -2)g}.\label{beta function of PL}
\end{equation}
For the noise phase to be included, we utilize the Gaussian function with zero mean as $\langle e^{\iota n\phi_i(t)}\rangle = e^{\eta(\tau)}$ where $\eta(\tau)=-\frac{1}{2}n^2\beta(\tau)$. To evaluate the noisy effects in the case of COM, BIP and TRI configurations, we take the average of the final density matrix given in Eq.\eqref{final density matrix} over the PL phase as \cite{Buscemi}:
\begin{equation}
\rho_{COM}(\tau)=\langle U(\phi_1,t)\rho_o U(\phi_1,t)^{\dagger} \rangle_{{\eta_1(\tau)}},\label{common environment}
\end{equation}
Where $ \phi_1(t)=\phi_2(t)=\phi_3(t) $ for an environment having a single source of noise. For the BIP configuration with two noise sources, the corresponding final density matrix can be obtained by \cite{Buscemi}:
\begin{equation}
\rho_{BIP}(\tau)=\langle \langle  U(\phi_1,\phi_3,t)\rho_o  U(\phi_1,\phi_3, t)^{\dagger} \rangle_{\eta_1(\tau)}\rangle_{\eta_3(\tau)},\label{mixed environment}
\end{equation}
here, we set $ \phi_1(t)=\phi_2(t) $. For the TRI configuration, the final density matrix of the three-qubit state reads as \cite{Buscemi}:
\begin{equation}
\rho_{TRI}(\tau)=\langle\langle \langle  U(\phi_1,\phi_2,\phi_3,t)\rho_o  U(\phi_1,\phi_2,\phi_3, t)^{\dagger}  \rangle_{\eta_1(\tau)} \rangle_{\eta_2(\tau)} \rangle_{\eta_3(\tau)}, \label{independent environment}
\end{equation}
where $ \langle \Psi \rangle $ represents the average over the possible values of the noise parameters over the state $ \Psi $.
\section{Non-local correlation and coherence measurement}
\subsection{Quantum negativity}
Quantum negativity is demonstrated to be an entanglement monotone and one of the most reliable entanglement measures. For tripartite quantum systems, negativity is computed by taking the geometric mean of the bipartite negativities of any possible system bipartition, and is written as:
\begin{align}
N(t)=\sqrt[3]{(N_{a|bc} N_{b|ac} N_{c|ab})},\label{Negativity}
\end{align}
where the negativity for the subsystem $a$ and joint system {bc} is defined as $N_{a|bc}=\sum_i|\lambda_i(\rho^{T_a})|-1 $. After performing the partial transpose regarding subsystem $a$, $\lambda_i(\rho^{T_a})$ is the $i^{th}$ eigenvalue of the density matrix of the three qubit states. If the density matrix $\rho^{T_a}$ has atleast one negative eigenvalue, then the corresponding negativity for the bipartition of the system is written as $\rho^{T_a}=2\hbox{max}\{0,-\lambda_{min} \}$ where, $\lambda_{min} $ is the smallest eigenvalue.
\subsection{Entanglement Witness Operation}
Entanglement witness is one of the operational method which distinguishes genuine entanglement from the bi-separability \cite{Li}. The defined entanglement witnesses for tripartite qubit states are defined as $\mathcal{E}_{1}=\frac{1}{2}I-\rho_{o} $, $\mathcal{E}_{2}=\frac{2}{3}I-\rho_{o} $, and $\mathcal{E}_{3}=\frac{3}{4}I-\rho_{o}$ \cite{Tchoffo}. Here, $I$ is the identity matrix and $\rho_o$ is the initial density matrix of the system. Entanglement witness can be computed by taking the negative trace of the time evolved state $\rho({\Delta,t})$ as \cite{Buscemi}:
\begin{equation}
EW(\tau)=-\hbox{Tr}[\mathcal{E}_{i}\rho({\Delta,t})]\label{EW operation}
\end{equation}
where $\mathcal{E}_{i}$ represent the corresponding entanglement witness. The outcome $\hbox{Tr}[\mathcal{E}_{i}\rho({\Delta,t})]< 0$ ensures the strong entanglement regime for the system and for $EW(\tau)=0$, the state will be assumed separable. However, note that the $\hbox{Tr}[\mathcal{E}_{i}\rho({\Delta,t})]>0$ does not ensure the absence of entanglement, especially for multiparty entanglement. This operation will be adopted to distinguish tripartite $GHZ$-class entangled states from the separable ones.
\subsection{Purity}
Purity operation estimtes the pureness and coherence of a quantum system \cite{Islam,pu1}. This measure for the time evolved density ensemble $\rho(\Delta,t)$ is given by:
\begin{equation}
P(t) =\hbox{Tr}[\rho(\Delta,t)]^{2}\label{purity}
\end{equation}
where, for an $n$-dimensional state, the purity ranges as $ \frac{1}{n}\leq P(t) \leq 1 $. For the lower bound $\frac{1}{n}$ of the purity, the state is completely mixed and decoherent, while for the upper bound, the state is pure and coherent.
\subsection{Entropy}
Entropy estimates the degree of environmental disorder and coherence loss in a quantum state \cite{Tolkunov}. The coherence loss results from the inevitable interaction of the quantum systems with the surrounding environments. We use von Neumann's entropy approach to estimate the loss of coherence and information in the three non-interacting qubits. Entropy for a time evolved density matrix $\rho(\Delta,t)$ reads as \cite{Maziero}:
\begin{equation}
EN(t)=-\hbox{Tr}[\rho(\Delta,t) \log \rho(\Delta,t)].\label{decoherence}
\end{equation}
Here, the state to be coherent with no information loss, $EN(t)=0$ while any other value of the measure will represent the corresponding coherence and information decay.
\section{Results and discussion}\label{Results and discussion}
The analytical simulations from the time evolved density matrix defined in Eq.\eqref{final density matrix} for the dynamics of the GHZ-like state are presented in this section. Using EW(t), P(t) and EN(t) measures from Eqs.\eqref{EW operation}, \eqref{Negativity}, \eqref{purity} and \eqref{decoherence}, we investigated the entanglement, coherence and information preservation by the current tripartite state.
\subsection{Characterization of isolated qubit-environment interaction and the related non-local correlation dynamics and protection in the local fields}
This section investigates the behavior and potential of the current stochastic fields to protect non-local correlation. We use entanglement witness as a function of $t$ to detect entanglement and separability in three maximally entangled non-interacting qubits. Eq.\eqref{final density matrix} is used to investigate the dynamical behaviour of the GHZ-like state before superimposing the noise phase over the joint phase of the system and local environments. In the current situation, the analytical results of the entanglement witness are as under (as shown in Figs.\ref{local field dynamics} and \ref{local field dynamics with p=0,1}):
\begin{figure}[ht]
\includegraphics[width=14cm, height=11cm]{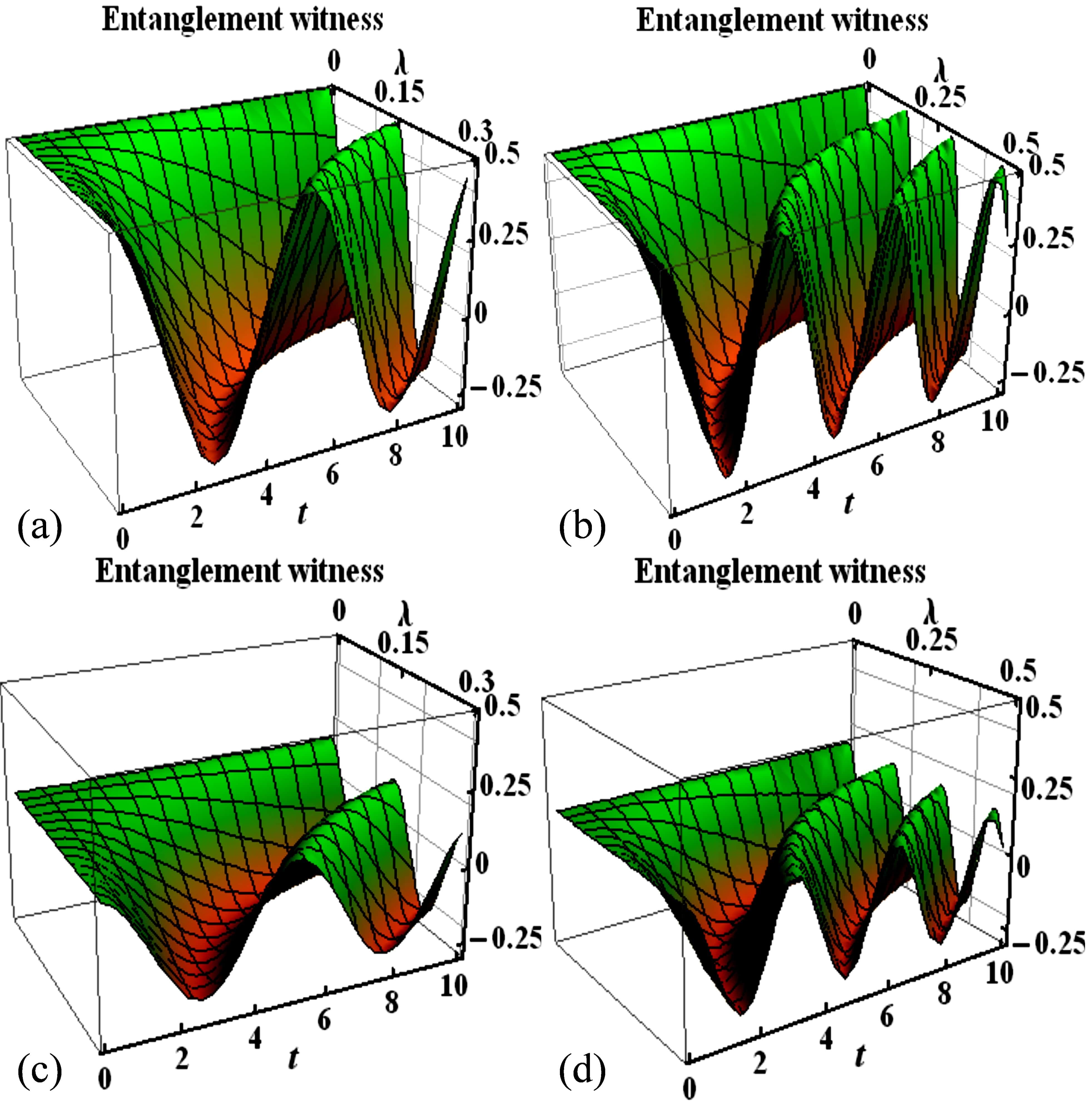}
\caption{Upper panel: Time evolution of the entanglement witness for the GHZ-like state with initial purity, $p=1$ subjected to local fields when $[\Delta_a,\Delta_b, \Delta_c] = 1$ with $\lambda=0.3$ (a) and $0.5$ (b). Bottom panel: same as upper column but when initial purity, $p=0.7$. }\label{local field dynamics}
\end{figure}
\begin{align}
EW_{COM}(t)=&\frac{1}{8} p (-4+5 p+3 p \text{Cos}[4 t \text{$\Delta $a} \lambda ]),\\
EW_{BIP}(t)=&\frac{1}{8} p (-4+3 p+p \text{Cos}[4 t \text{$\Delta $a} \lambda ]+2 p (\text{Cos}[2 t (\text{$\Delta $a}-\text{$\Delta $b}) \lambda ]+\text{Cos}[2 t (\text{$\Delta $a}+\text{$\Delta $b}) \lambda ])),\\
\begin{split}
EW_{TRI}(t)=&\frac{1}{8} p (2 (-2+p)+p (\text{Cos}[2 t (\text{$\Delta $a}-\text{$\Delta $b}) \lambda ]+\text{Cos}[2 t (\text{$\Delta $a}+\text{$\Delta $b}) \lambda ]+\\&
\text{Cos}[2 t (\text{$\Delta $a}-\text{$\Delta $c}) \lambda ]+\text{Cos}[2 t (\text{$\Delta $b}-\text{$\Delta $c}) \lambda ]+\text{Cos}[2 t (\text{$\Delta $a}+\text{$\Delta $c}) \lambda ]\\& +\text{Cos}[2 t (\text{$\Delta $b}+\text{$\Delta $c}) \lambda ])).
\end{split}
\end{align}
Fig.\ref{local field dynamics} evaluates the dynamical behaviour of GHZ-like state along with the detectable non-local correlation in local random fields. We have set $\Delta_a=\Delta_b=\Delta_c$ for COM, BIP and TRI environments for $\lambda=0.3$ and $0.5$. The frequency of oscillations varies significantly between the two $\lambda$ values and rapidly increases when this parameter is increased. No damping is observed because no noise phase is superimposed over the common phase of the system and environment. As a result, noise-free classical environments are a valuable resource in quantum information processing. Although, a fully noise-free environment would be ideal, it is unattainable; nonetheless, noises can be minimized to some extent. As a result, it's vital to recognize the noises, as well as their contents and associated disorders. This is vital when comparing the primary role of a pure classical environment with no distractions to that of a noisy environment. We will compare the system's time evolution in noiseless and noisy contexts using the current description, allowing us to estimate non-local correlation and coherence losses. For the existing classical fields, we included the initial purity factor ($p$) in the current results for $p=1$ (upper panel) and $0.7$ (bottom panel). The amplitude of fluctuations in the current results entirely depends on the system's initial purity factor.\\
\begin{figure}[ht]
\includegraphics[width=14cm, height=6cm]{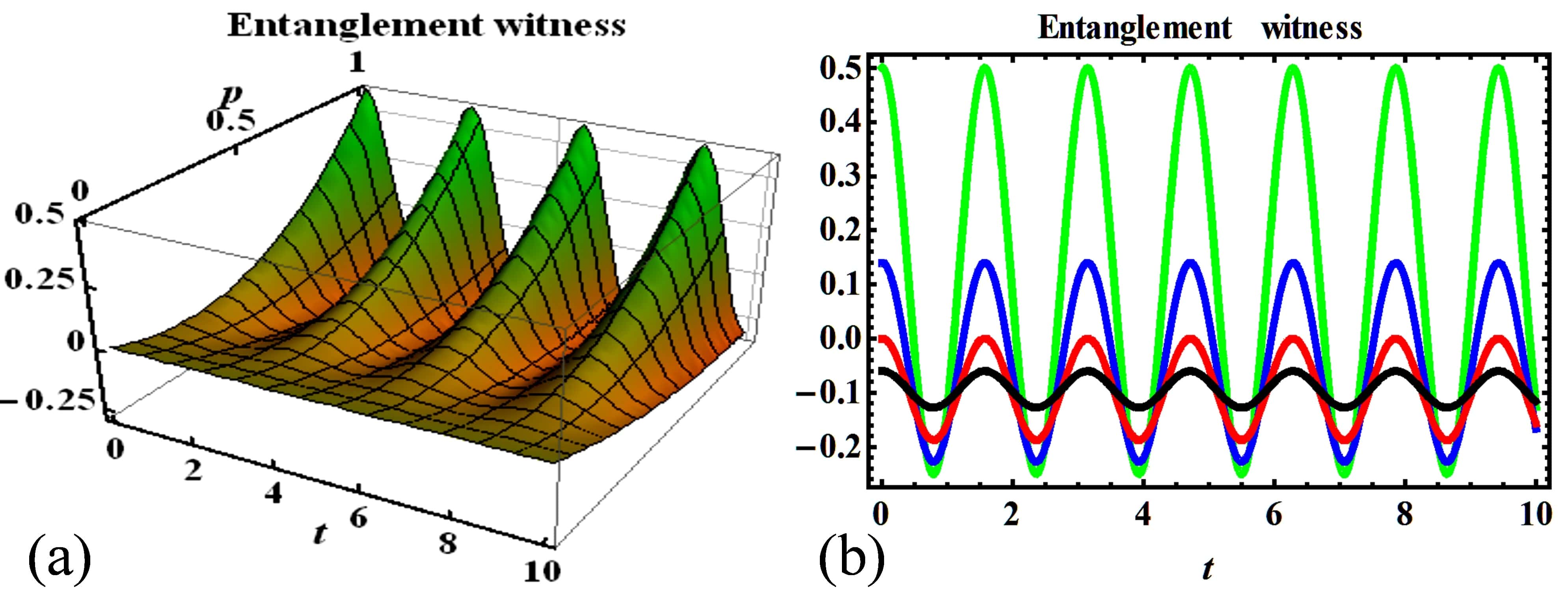}
\caption{Fig.3(a): Time evolution of the entanglement witness for the GHZ-like state with initial purity, $0\leq p \leq 1$ subjected to local fields when $[\Delta_a,\Delta_b, \Delta_c] = 1$, $\lambda=0.7$. Fig.3(b): Same as Fig.3(a) but when $p=1$ (green), $0.7$ (blue), $0.5$ (red) and $0.2$ (black) with $\lambda=1$.}\label{local field dynamics with p=0,1}
\end{figure}
In Fig.\ref{local field dynamics with p=0,1}, the role of the initial purity of the GHZ-class state when coupled with classical environments is evaluated. For different values of $p$, the dynamical outlook is distinguishable from the current results. In Fig.3(a), the amplitude of the system decreases proportionally to the decreasing value of $p$. The amplitude of fluctuation increases directly as $p$ increases in Fig.3(b) for initial purity, $p=1$ (green lines) and $0.2$ (black lines). Most notably, we realized that the GHZ-class state is completely separable at $p \leq 0.5$. This means that in the current context, the GHZ-class state will be entangled for the purity range of $0.5 \leq p \leq 1$, with maximum detectable entangled occurring at the upper bound. This is consistent with the definition of the ground state of GHZ-class system given in Eq.\eqref{initial denisty matrix state}. It is important to note that the initial purity of the system has no effect on the system's fluctuation rate and only regulates the associated amplitudes. In classical environments, entanglement sudden death (ESD) and entanglement sudden birth (ESB) revivals have been strongly observed. Significantly, the slopes depict the dynamics of the $EW(t)$ in negative regimes, implying the state's separability. However, the state re-enters the strong entanglement regimes, i.e. positive $EW(t)$ regimes. It is interesting to note that the entanglement amplitude is greater than the corresponding separability, ensuring that the state is not completely disentangled at any point in the dynamical map. This indicates that the environments have a strong back-action effect, converting resource states to free states and free states back to resource states.
\subsection{Dynamical map of negativity under PL noise}
\begin{figure}[H]
\includegraphics[width=16.5cm, height=11cm]{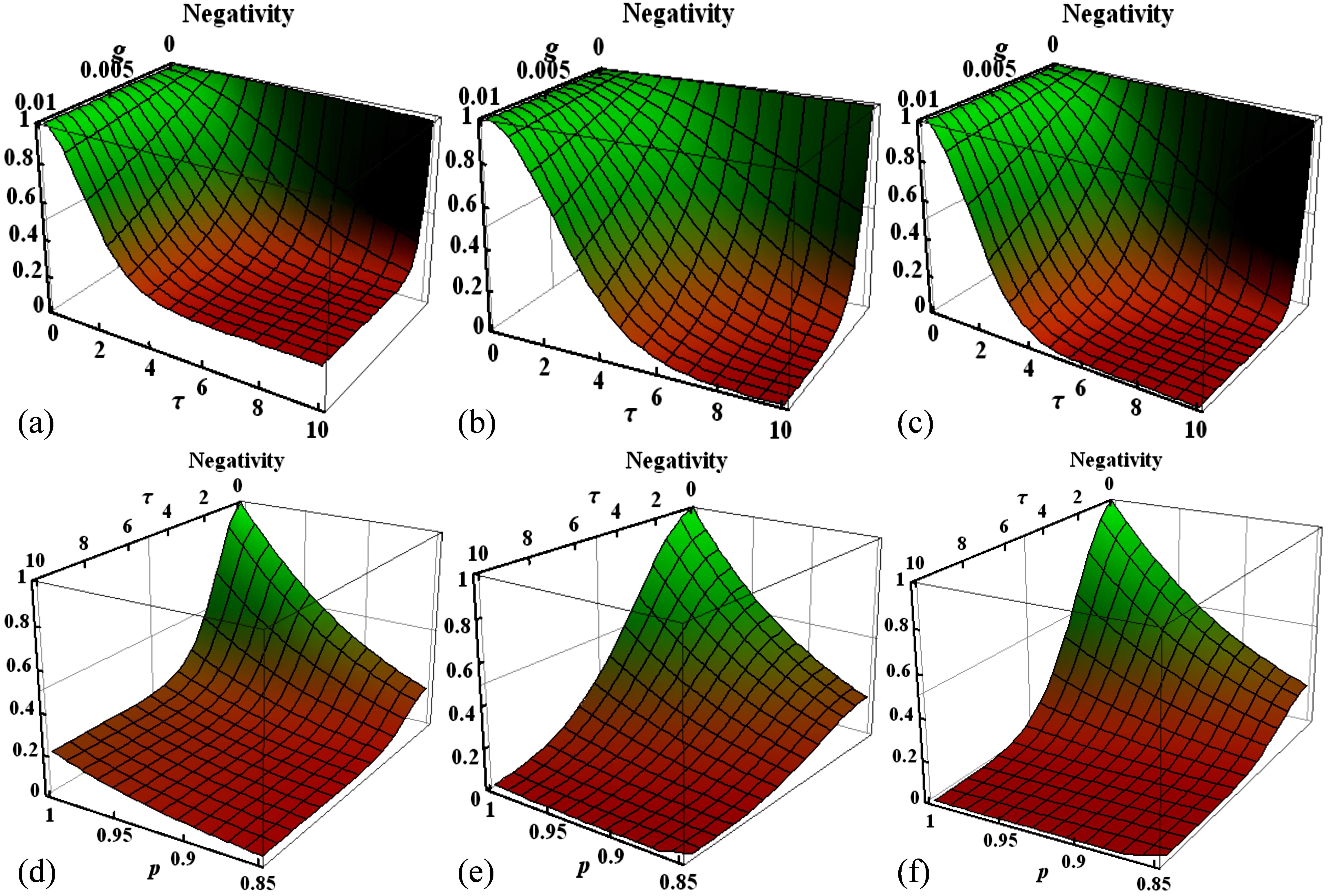}
\caption{Upper panel: Time evolution of the negativity as a function of $\tau$ for the GHZ-like state with initial purity, $p=1$ when subjected to common (a), bipartite (b) and tripartite (c) configurations under the effects of PL noise with the following parameter: $g=10^{-2}$ and $\alpha=3$. Bottom panel: Same as upper panel, but for the negativity as a function of $p$.}\label{negativity 3d graphs}
\end{figure}

Entanglement measurement through $N(\tau)$ is explored by applying Eq.\eqref{Negativity} for the dynamics of a GHZ-like state in the presence of PL noise in this section. We used Eqs.\eqref{common environment}, \eqref{mixed environment}, and \eqref{independent environment} for the COM, BIP, and TRI configurations. Note that we have set the initial purity of the system to $1$ and in the range $0.85\leq p\leq 1$. The corresponding results are followed as (also shown in Fig.\ref{negativity 3d graphs}):\\
\begin{align}
N_{COM}(\tau)=&\left(-1+\frac{\lvert p\rvert}{2}+\frac{1}{4}( \sum_{i=1} ^2 \mathcal{A}_i p)+ \mathcal{A}_3 P\right)^x,\\
N_{BIP}(\tau)=&\left(-1+\frac{\lvert p \rvert}{2}+\mathcal{A}_4+\frac{1}{4} \mathcal{A}_5 p]+\frac{1}{4} \mathcal{A}_6 p]) (\mathcal{A}_7p)+\frac{1}{2} \mathcal{A}_8 p+\frac{1}{8} \sum_{i=9}^{12}\mathcal{A}_i\right)^y,\\
N_{TRI}(\tau)=&\left(-1+\mathcal{A}_{13} \vert p \vert+\frac{1}{2} \mathcal{A}_{14}p]+\frac{1}{2} \mathcal{A}_{15} p]\right)^z.
\end{align}
Fig.\ref{negativity 3d graphs} analyses the dynamics of quantum negativity for GHZ-class state, initially prepared in the state $\rho_o$ of ground purity $1$ (upper panel) and in the range $0.85 \leq p \leq 1$ (bottom panel) when coupled to COM, BIP and TRI configurations. By comparing Figs.\ref{local field dynamics} and \ref{negativity 3d graphs}, we conclude that the classical environment when exploited with noise becomes inefficient and suffers significant loss of entanglement. We find the relative entanglement dynamics and associated preservation varied significantly over time in environments with varying compositions. In COM configuration, the GHZ-class state remained entangled, but it became fully separable in BIP and TRI configurations. COM configurations are thus valuable resources in the presence of PL noise for avoiding the conversion of GHZ-class states to free states. In classical environments, the PL noisy effects are like the Ornstein Uhlenbeck and fractional Gaussian noisy effects observed in \cite{Kenfack, Rossi, ATTA-OU, ATTA-PL}. The entanglement dissipated faster as the value of $g$ increased. As a result, it was discovered that parameter $g$ has a negative impact on the memory properties of the environments, resulting in entanglement degradation. The observed decay has a completely monotonous time function, with no revivals. Therefore, classical environments highly suppressed the ESD and ESB revivals. This demonstrates that the environment has no reversible effects and that the information loss is irreversible. The PL noisy effects can be related to the quantum to signal-noise ratio results obtained in \cite{Benedetti}. The initial purity of the GHZ-class quantum system in the ground state $\rho_o$ was discovered to be crucial in altering the amplitude of the entanglement. At unit purity, the degree of entanglement is greater than at other values. Hence, preparing GHZ-like states with initial unit purity is more helpful for the preservation of quantum information and non-local correlations. We discover it in contrast to the disentanglement of three GHZ-class qubits in the purity range $\frac{1}{4} \leq p \leq 1$ due to Ornstein Uhlenbeck noise in the COM configuration \cite{Kenfack}. We find that the asymptotic saturation values of the $N(\tau)$ are zero in the BIP and TRI configurations.
\begin{table}[b]
\begin{tabular}{lclclclclcl}
\rowcolor{green!10}
Configuration &P-factor &  E-level &E-time 	 & Previous results \\ \\
\rowcolor{blue!3}COM	&1    & 0.2		& indefinite &Using concurrence, the bipartite states \\\rowcolor{blue!5}
	&0.85 &	0		&3 			 &in COM configuration under PL noise \\\rowcolor{blue!5}
	&	  &			& 			 &becomes separable \cite{ATTA-PLFG}.\\\rowcolor{pink!5}
BIP &  	  & 0		& 8			 & Using negativity, the three level system under\\\rowcolor{pink!5}
	&0.85 &	0		&7 			 & RTN noise remained entangled for longer \\\rowcolor{pink!5}
	&	  &			& 			 & time in BIP configuration \cite{Fabrice}.\\\rowcolor{blue!5}
TRI &1 	  & 0		& 5.5		 & Using entanglement witness, the tripartite state\\\rowcolor{blue!5}
	&0.85 &	0		&7 			 & under the Gaussian noise in TRI scheme becomes\\\rowcolor{blue!5}
	&	  &			& 			 &becomes disentangled in a short time \cite{ATTA-GN}.\\	
\end{tabular}
\caption{Shows the quantitaive study of the entanglement preservation versus initial purity of the GHZ-like state using negativty under PL noise along with the previous investigations. Here, P-factor represent the initial purity, E-level and E-time indicates the entanglement level and entanglement time obtained through negativity in Fig.\ref{negativity 3d graphs}. Note that, the values are obtained from the final saturation levels of the entanglement.}\label{table1}
\end{table}
This is similar to $N(\tau)$ results obtained under the Ornstein Uhlenbeck noise and in the Markov regimes of random telegraphs and coloured noises, but with a different dynamical map, \cite{Kenfack, Tchoffo}. Because of the entanglement generation between the system and the environment, we conclude that the initially prepared pure GHZ-class states transform into mixed states under PL noise except, in the case of COM configuration at $p=1$. The dynamics of quantum discord obtained under various non-Gaussian noises differ completely from the current time evolution of the $N(\tau)$ under PL noise \cite{Tchoffo, Kenfack-qbqt, Javed}. Aside from that, we discovered that the GHZ-class state remained more stable, entangled, and coherent than the two-qubit maximal entangled state, as seen in \cite{ATTA-PLFG}, under PL noise.
\subsection{Dynamical map of entanglement witness under PL noise}
This section explores the entanglement through $EW(\tau)$ by using Eq.\eqref{EW operation} for the dynamics of GHZ-like state when $p=1$ or $0.5 \leq p \leq 1$ under the presence of PL noise. For the COM, BIP and TRI configurations, we employed Eqs.\eqref{common environment}, \eqref{mixed environment} and \eqref{independent environment}. The corresponding results are followed as (also shown in Fig.\ref{ewo 3d graphs}):
\begin{figure}[ht]
\includegraphics[width=16.5cm, height=11cm]{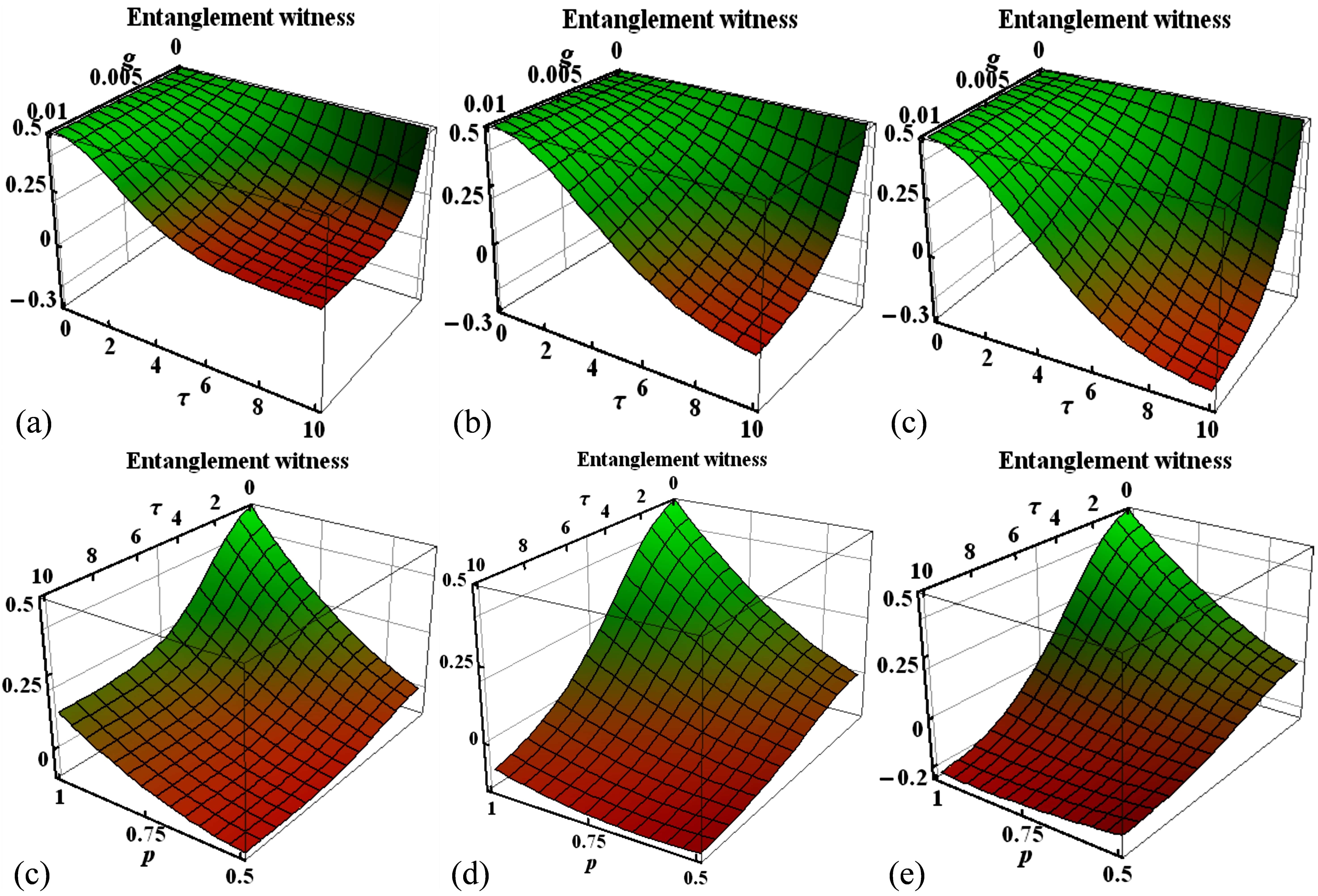}
\caption{Upper panel: Time evolution of the entanglement witness as a function of $\tau$ for the GHZ-like state with initial purity, $p=1$ when subjected to common (a), bipartite (b) and tripartite (c) configurations under the effects of PL noise with the following parameter: $g=10^{-2}$ and $\alpha=3$. Bottom panel: Same as upper pannel but for the entanglement witness as a function of $p$.}\label{ewo 3d graphs}
\end{figure}
\begin{align}
\mathcal{EW}\rho_{COM}(\tau)=&\frac{1}{8} p \left(-4+(5+3 \Omega_1) p\right),\\
\mathcal{EW}\rho_{BIP}(\tau)=&\frac{1}{8} \Omega_2 p \left(4 p+\Omega_3 p+\Omega2(-4+3 p)\right),\\
\mathcal{EW}\rho_{TRI}(\tau)=&\frac{1}{4} p \left(-2+p+3 \Omega_2 p\right).
\end{align}
Fig.\ref{ewo 3d graphs} evaluates the dynamics of $EW(\tau)$ for the GHZ-like state when exposed to classical random fields in the presence of PL noise. $EW(\tau)$ remained a decreasing function of entanglement that demonstrates the dominant distortion nature of PL noise in which the originally encoded entanglement rapidly decays. Without the ESD and ESB events, the qualitative degradation viewpoint is completely monotone, with no revivals. This means that, except for the three qubits in the GHZ-class in COM configuration, the three-qubit states become free and do not revert to resource states. When exposed to a single PL noise source for an indefinite time while remaining partially entangled, the GHZ state has been discovered to be a good resource. All the results obtained in Fig.\ref{ewo 3d graphs} strongly disagree with those shown in Fig.\ref{local field dynamics}. This suggests strong contrasting behvaiour between the isolated and non-isolated qubit-environment interactions. Besides, according to Ref.\cite{Buscemi, Tchoffo, Kenfack-qbqt}, the current findings contradict the tripartite quantum system dynamics observed under coloured and random telegraph noises, where significant ESD and ESB revivals were observed. Entanglement of bipartite and tripartite states, on the other hand, match the current dynamical picture under fractional Gaussian and Ornstein Uhlenbeck noise, as cited in \cite{Kenfack, Rossi, ATTA-OU, ATTA-PLFG}.
\begin{table}[b]
\begin{tabular}{lclclclclcl}
\rowcolor{green!10}
Configuration &P-factor &  E-level &E-time 	 & Previous results \\ \\
\rowcolor{blue!3}
COM &1    & 0.2	& indefinite &Using entanglement witness, the bipartite state in\\\rowcolor{blue!3}
	&0.85 &	0	& 4			 &OCM configuration under PL noise becomes separ-\\\rowcolor{blue!3}
	&	  &		&   		 &able either in a long or short time, relatively \cite{ATTA-PLFG}.\\\rowcolor{black!0.5}
BIP &1    & -0.1& 5 		 &Using entanglement witness, the tripartite state in\\\rowcolor{black!0.5}
	&0.85 &	-0.1& 7 		 &BIP configuration under Gaussian noise becomes \\\rowcolor{black!0.5}
	&	  &-0.15&  			 &separable in a short time, comparatively \cite{ATTA-GN}.\\\rowcolor{blue!3}
TRI &1    &-0.2	& 4 		 &Using negativity and GQD, GHZ and W-type states  \\\rowcolor{blue!3}
	&0.85 &-0.1	& 5			 &have been detected in entanglement regime with\\\rowcolor{blue!3}
	&	  &-0.2 &			  &different purity factor than the current one \cite{Kenfack-qbqt}.\\	
\end{tabular}
\caption{Shows the quantitaive study of the entanglement preservation versus initial purity of the GHZ-like state using entanglement witness under PL noise along with the previous investigations. Here, P-factor represent the initial purity, E-level and E-time indicates the entanglement level and entanglement time obtained through entanglement witness in Fig.\ref{ewo 3d graphs}. Besides, GQD is the geometric quantum discord used to quantify quantum correlations in \cite{Kenfack-qbqt}. Note that the values are obtained from the final saturation levels (i.e. when slopes reaches $0$ in the case of entanglement witness) of the entanglement.}\label{table2}
\end{table}
Depending on the configuration, entanglement and coherence protection across different environments varies quantitatively. The GHZ-like state was discovered to be more robust for entanglement retention in the COM configuration and completely separable after a certain duration in both the BIP and TRI schemes. This can be traced back to the results obtained for tripartite and multipartite states in classical environments controlled by Gaussian type noises, as demonstrated in \cite{ATTA-PLFG, Kenfack, ATTA-OU}, but with different preservation thresholds. The current results demonstrate that the parameter $g$ is decoherent in terms of entanglement protection. As seen, entanglement deterioration increases as $g$ increases. The Hurst index's role in fractional Gaussian noise is opposed to the current noisy role of the parameter $g$ \cite{Rossi}. The initial purity factor for GHZ-class states was encountered to regulate the amplitude of entanglement. The entanglement preservation interval is longer at $p=1$. When $p<1$, the asymptotic values of the $EW(\tau)$ obtained are zero for all input parameter values, except in the case of COM configuration between the range $0.5 \leq p \leq 1$. The results of $EW(\tau)$ agree well with those of $N(\tau)$, indicating that there are strong links between them and reliability. The current $EW(\tau)$ results obtained under PL noise contradict those obtained under static noise for the same purity factor range studied in \cite{Kenfack-static(p)}. 
\subsection{Dynamical map of purity under PL noise}
\begin{figure}[b]
\includegraphics[width=16.5cm, height=11cm]{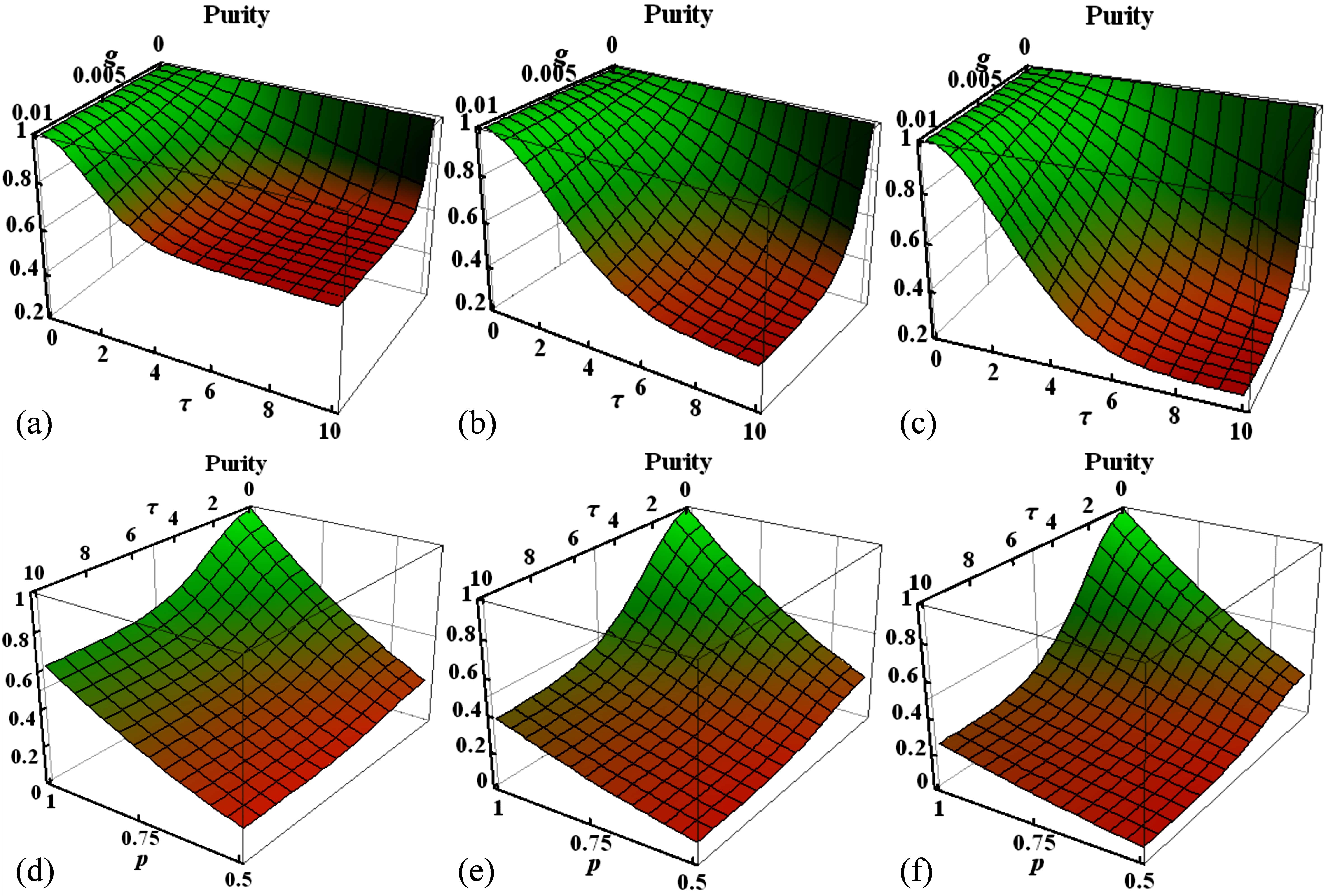}
\caption{Upper panel: Time evolution of the purity as a function of $\tau$ for the GHZ-like state with initial purity, $p=1$ when subjected to common (a), bipartite (b) and tripartite (c) configurations under the effects of PL noise with the following parameter: $g=10^{-2}$ and $\alpha=3$. Bottom panel: Same as upper panel, but for the purity as a function of $p$.}\label{purity 3d graphs}
\end{figure}
When $p=1$ or $0.5 \leq p \leq 1$, the numerical simulations for the time evolution of the $P(\tau)$ for the dynamics of the GHZ-like state under PL noise are presented using Eq.\eqref{purity}. By deploying Eqs.\eqref{common environment}, \eqref{mixed environment}, and \eqref{independent environment} for the COM, BIP and TRI configurations, the $P(\tau)$ gives the corresponding results as followed as (also sown in Fig.\ref{purity 3d graphs}):
\begin{align}
P_{COM}(\tau)=&\xi_1 \left(5+3 \varrho_1\right) p^2,\\
P_{BIP}(\tau)=&\xi_1 \varrho_2 \left(4+\Omega_2+3 \varrho_2\right) p^2,\\
P_{TRI}(\tau)=&\xi_2 \varrho_2 \left(3+\varrho_2\right) p^2.
\end{align}
Fig.\ref{purity 3d graphs} shows the time evolution of the $P(\tau)$ for GHZ-like state when subjected to classical fields under the presence of PL noise. $P(\tau)$ has been found to be a decreasing function of the coherence concerning $\tau$. The findings show that in the tripartite GHZ state, PL noise has a detrimental effect on coherence and purity preservation. In \cite{Benedetti}, comparable noisy effects related to the current noise for quantum signals to noise ratio were also presented. $P(\tau)$ shows monotonic decay for the current system under PL noise, which is consistent with $EW(\tau)$, implying that ESD and ESB revivals are not possible. This indicates that the degradation is permanent, and that the lost information cannot be retrieved to the system in all situations due to no back-action of the environments. The GHZ-like state stays coherent in the COM configuration, according to the $P(\tau)$ measure, which is consistent with previous $EW(\tau)$ and $N(\tau)$ findings. Thus, $P(\tau)$ offers a simple and reliable approach for distinguishing coherent and pure states from non-coherent and mixed states. When the $g$ value range is expanded, the decoherent character of PL noise becomes more apparent. The current results of monotonic degradation and detecting robustness in COM configuration with no coherence revivals have also been observed due to Ornstein Uhlenbeck noise in \cite{Kenfack, ATTA-OU}. Significant osscilatory events, however, have been discovered for the dynamics of the tripartite states under coloured noise, as presented in \cite{Buscemi, Kenfack-static(p)}.
\begin{table}[b]
\begin{tabular}{lclclclclcl}
\rowcolor{green!10}
Configuration &P-factor &  C-level &C-time 	 & Previous results \\ \\
\rowcolor{blue!3}
COM	&1   & 0.65	 & indefinite & Using negativity and QD, the qubit-qutrit state\\\rowcolor{blue!3}
	&0.5 &	0.2	 & 0 		  &	under dephasing and depolarizing noises becomes \\\rowcolor{blue!3}
	&	 &		 &			  &disentangled and decoherent in a short time \cite{Karpat}.\\\rowcolor{black!0.5}
BIP &1   & 0.4	 & 5		  & Different qutrit states were found very fragile to\\\rowcolor{black!0.5}
	&0.5 &0.15   & 0		  &preserve entanglement and coherence against the \\\rowcolor{black!0.5}
	&	 & 		 & 			  & stochastic dephasing \cite{Xiao}.\\	\rowcolor{blue!3}
TRI &1   & 0.25	 & 4		  & GHZ and W-type states have been detected with\\\rowcolor{blue!3}
	&0.5 &0.1	 & 0		  &	long preserved entanglement in TRI configuration  \\\rowcolor{blue!3}
	&	 &		 &			  & under RTN noise using negativity \cite{Buscemi}.\\
\end{tabular}
\caption{Shows the quantitaive study of the coherence preservation versus initial purity of the GHZ-like state using purity measure under PL noise along with the previous investigations. Here, P-factor represent the initial purity, C-level and C-time indicates the coehernce level and coherence time obtained through purity measure in Fig.\ref{purity 3d graphs}. Besides, QD is the quantum discord used to quantify quantum correlations in \cite{Kenfack-qbqt}. Note that the values are obtained from the final saturation levels (i.e. when slopes reaches $0.5$ in the case of purity measure) of the coherence.}\label{table3}
\end{table}
When $EW(\tau)$, $P(\tau)$, and E$N(\tau)$ are compared, it can be seen that the disentanglement rate in the three-qubit GHZ state is proportional to the increase in decoherence. In terms of entanglement and coherence protection, the current GHZ-like state outperforms the W-type states studied in \cite{Kenfack-static(p)} in classical environments. The W-type three-qubit states mentioned in \cite{Tchoffo, Kenfack-static(p)} become unentangled and decoherent in all environments, contradicting the current findings, particularly in the case of COM setup. When compared to current classical environments, Gaussian distributed environments exhibit high ESD and ESB revivals \cite{Kenfack-GDE}. The present results for the input parameter p correspond well with those of $N(\tau)$ and EW (t). $P(\tau)$ shows that at $p=1$ and the lower bound of $g$, the maximum entanglement and coherence with non-zero asymptotic values occur. For $p \leq 0.75$, the state is entirely separable and decoherent, and it approaches the minimal bound of the $P(\tau)$ measure (i.e. $0.5$). As a result, the state in the COM configuration remains entangled in the range $0.5 \leq p \leq 1$ indefinitely. In contrast, the state in BIP and TRI configurations remains entangled and coherent for limited intervals. This also considers the fact that when the initial purity factor is reduced, the state becomes easily entangled with its surroundings, resulting in information loss. Matching results are obtained for the tripartite states under Gaussian and non-Gaussian noise, for example, in \cite{Kenfack, Kenfack-static(p)}, with varying preservation intervals and p ranges.
\subsection{Dynamical map of entropy under PL noise}
\begin{figure}[b]
\centering
\includegraphics[width=16.5cm, height=11cm]{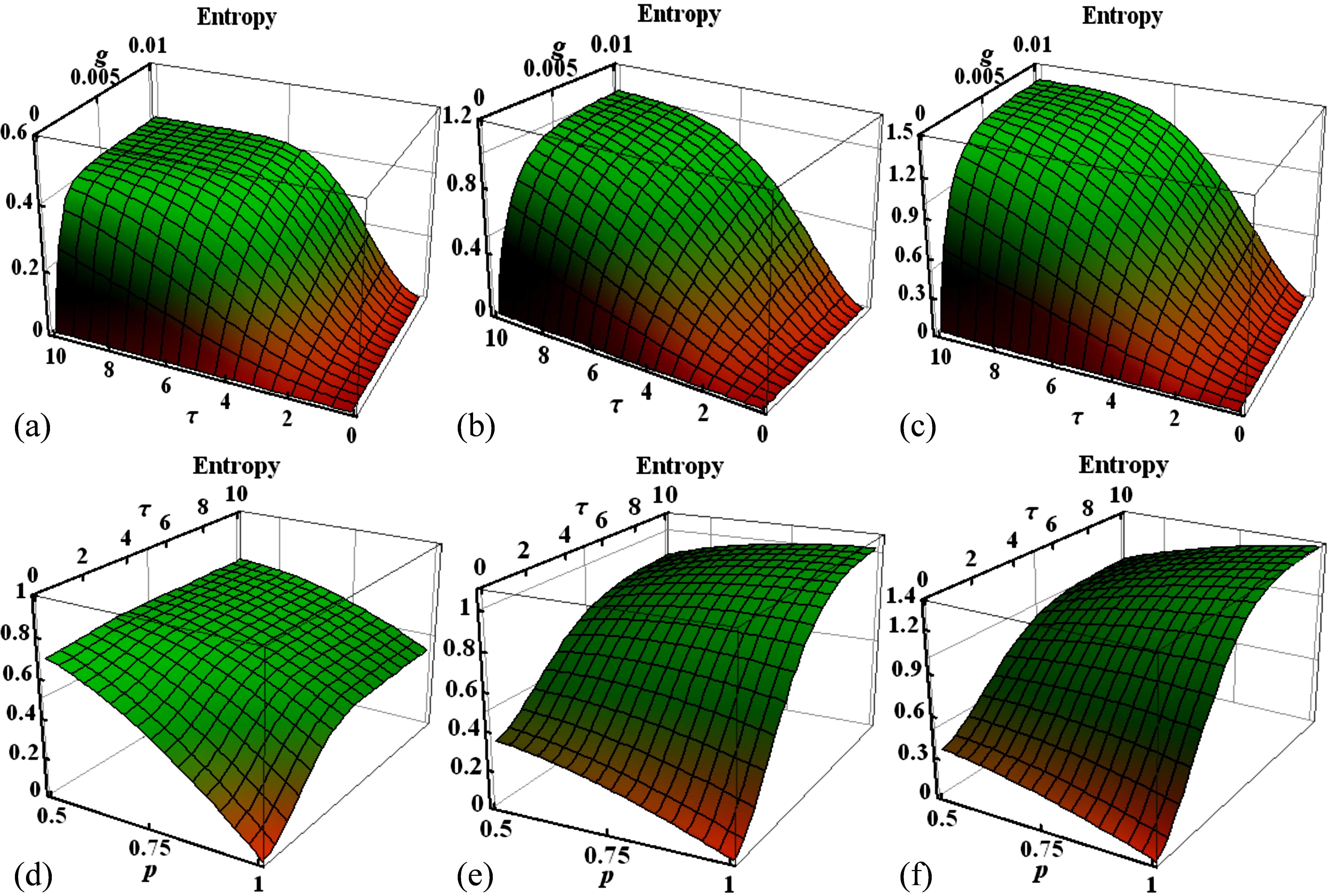}
\caption{Upper panel: Time evolution of the entropy as a function of $\tau$ for the GHZ-like state with initial purity, $p=1$ when subjected to common (a), bipartite (b) and tripartite (c) configurations under the effects of PL noise with the following parameter: $g=10^{-2}$ and $\alpha=3$. Bottom panel: Same as upper pannel but for the entropy as a function of $p$.}\label{decoherence 3d graphs}
\end{figure}
The explicit results are reported using Eq.\eqref{decoherence} for the time evolution of the entropy for the GHZ-like state under PL noise with initial purity, $p=1$ or $0.5 \leq p \leq 1$. $EN(\tau)$ produces the following findings (also displayed in Fig.\ref{decoherence 3d graphs}) by employing Eqs.\eqref{common environment}, \eqref{mixed environment}, and \eqref{independent environment} for COM, BIP and TRI configurations:
\begin{align}
EN_{COM}(\tau)=&-\mu_1 (2 p+\chi_1) \log\left[2\mu_1 (2 p+\chi_1)\right],\\
EN_{BIP}(\tau)=&\frac{\mu_1}{2}\chi_2 \log \left[\frac{\mu_1}{2} \xi_2\right]-\frac{\mu_1}{4} \chi_3 \log\left[\frac{\mu_1}{4} \chi_3\right]-\frac{\mu_1}{4} \chi_4 \log\left[\frac{\mu_1}{4} \chi_4\right],\\
EN_{TRI}(\tau)=&-\frac{\mu_1}{2} \chi_5 \log \left[\frac{\mu_1}{2}\chi_5 \right]-\mu_2 \chi_6 \log\left[\mu_2 \chi_6\right].
\end{align}
The dynamics of entropy for the GHZ-like state under PL noise is shown in Fig.\ref{decoherence 3d graphs}. The current dynamical process is investigated using three alternative configurations: COM, BIP, and TRI. Unlike the $N(\tau)$, $EW(\tau)$, and $P(\tau)$, we noticed that the $EN(\tau)$ is an increasing function of coherence decay with respect to $\tau$. According to $EN(\tau)$, the coherence decay is pure exponential, and no coherence revivals have been observed. This type of decay indicates that the lost data is not being returned to the system from the environment. Hence, information lost because of the PL noise is irreversible and cannot be recovered using any system-environment coupling strategy. $EN(\tau)$ also implies that when coupled to a COM configuration, the GHZ-like state will remain more dominant for coherence preservation. As can be seen, the amount of decoherence for the GHZ-like state under a single noise is the smallest. Furthermore, the observed decay appears to increase with increasing $g$ value. Thus, by properly fixing $g$, one can control the decoherent nature for optimal results. Matching results with the current are observed in the tripartite states under Ornstein Uhlenbeck and fractional Gaussian noise \cite{Kenfack, Rossi}. Surprisingly, the decoherence measurement for the same three states under static and RTN noise revealed a revival decay rather than a monotonic decay \cite{Tchoffo, Lionel, Kenfack-static(p)}. The current role of classical environments controlled by Gaussian PL noise is completely contrary to the Gaussian distributed environments investigated in \cite{Kenfack-GDE}. The findings are consistent for the entanglement and coherence preservation measured by $N(\tau)$, $EW(\tau)$ and $P(\tau)$ in response to different input parameter p ranges. We discover that the minimum decoherence effects are only observed at the upper bound of the input parameter p.
\begin{table}[ht]
\begin{tabular}{lclclclclcl}
\rowcolor{green!10}
Configuration &P-factor &  C-level &C-time 	 & Previous results \\ \\
\rowcolor{blue!3}
COM	 &1	  & 0.44	& 7 	&Bipartite mixture states have less preserved\\\rowcolor{blue!3}
	&0.5  &	0.7		& 3 	&quantum correlations against varied P-factors \\\rowcolor{blue!3}
	&	  &			&	    &in COM configuration with static noise \cite{Shamirzaie}.\\\rowcolor{black!0.5}
BIP &1 	  & 1.1		& 8.5   &GHZ and W-type states faced faster decay of\\\rowcolor{black!0.5}
	&0.5  &	0.82	&10 	&entanglement and coherence in BIP configur- \\\rowcolor{black!0.5}
	&	  &			& 		&ation with static noises \cite{Tchoffo}.\\\rowcolor{blue!5}
TRI &1 	  & 1.4		& 9 	&Sudden death of entanglement and coherence\\\rowcolor{blue!5}
	&0.5  &	0.92	&7 		&are observed for bipartite states under \\\rowcolor{blue!5}
	&	  &			&	    &classical dephasing noises \cite{Yu}.\\	
\end{tabular}
\caption{Shows the quantitaive study of the coherence preservation and related disorder level versus initial purity of the GHZ-like state using entropy under PL noise along with the previous investigations. Here, P-factor represent the initial purity, C-level and SN-time indicates the coherence level and saturation time obtained through von Neumann entropy in Fig.\ref{decoherence 3d graphs}. Note that the values are obtained from the final saturation levels of the coherence.}\label{table4}
\end{table}
\subsection{Non-local correlation and coherence retention potential of the local environments}
Here, we provide brief results for the dynamics of the $N(\tau)$, $EW(\tau)$, $P(\tau)$ and $EN(\tau)$ for the GHZ-like state using Eqs.\eqref{negativity 3d graphs}, \eqref{EW operation}, \eqref{purity} and \eqref{decoherence}. In this section, we explore briefly the prolong time evolution process for the wide spectrum of the PL noise. We also include the characterization of the entanglement and coherence dynamics counter to various values of the $g$, to exploit the noisy effects of PL noise in COM, BIP and TRI configuration.\\ 
\begin{figure}[ht]
\includegraphics[width=16.5cm, height=4.2cm]{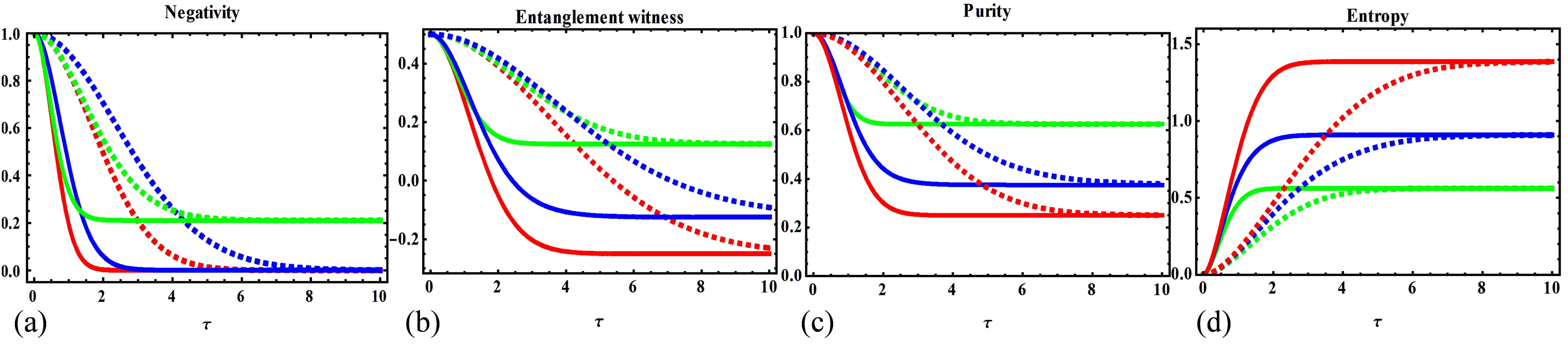}
\caption{Time evolution of negativty (a), entanglement witness (b), purity (c) and entropy (d) for the GHZ-like state when subjected to COM (green), BIP (blue), TRI (red) with the following parameter: $g=10^{-2}$ (dashed), $10^{-1}$ (dashed) $\alpha=3$ and $p=1$.}\label{GHZ state under PLN}
\end{figure}
Fig.\ref{GHZ state under PLN} shows the dynamics of the GHZ-like state when coupled to common (green), bipartite (blue), and tripartite (red) configuration. The $N(\tau)$, $EW(\tau)$, and $P(\tau)$ measures, all were found decreasing functions of entanglement and coherence, whereas E$N(\tau)$ remained rising function of coherence decay. This investigation explains the destructive nature of the PL noise against the retention of non-local correlations and coherence. The statement is accompanied by a comparison of the final to the initial encoded entanglement and coherence in the state. In the case of environments, the decay encountered is much smaller in COM configuration, followed by BIP, and corresponds to the result given in \cite{Kenfack}. It is worth noting that the GHZ-like state preserves entanglement and coherence in the COM configuration, which contradicts the $N(\tau)$ and quantum discord conclusions for maximally entangled two qubits and mixture states, for example examined in \cite{Rossi, Benedetti-color noise}. Although the qualitative decay behaviour varies depending on the environment and parameter values involved, it is fully monotonic, with no entanglement revivals. Thus, the GHZ-like state is devoid of ESD and the ESD phenomena besides the coherence revivals. This interprets the irreversible loss of entanglement and coherence owing to the unavailability of a back-flow mechanism between the system and environment. This also evaluates the fact that, in the current Gaussian noise assisted environments, converting the free states back into the required resource states with the original encoded information is not possible. This contradicts the quantum dynamics observed in \cite{Kenfack-GDE} under Gaussian distributed classical environments characterized by Ornstein Uhlenbeck noise. By comparing previously studied tripartite, bipartite pure and hybrid quantum states, it is possible to conclude that the existence of revivals entirely depends on the type of noise and environment involved, as the same states are found to have revivals against Gaussian, static, and coloured noises given in \cite{Tchoffo, Lionel, Kenfack-static(p), Kenfack-GDE, Kenfack-qbqt}. Aside from that, the revivals also depend on the special properties of the environmental noise, such as the Markov and non-Markov characteristics. The former statements can be seen by comparing the various dynamical studies provided in \cite{Kenfack, Rossi, Zhang, Tchoffo, Buscemi, Lionel, Javed, Benedetti-color noise, Yu}. As $g$ increases, the slopes of noise parameters are observed to shift from the green to the black end. This implies that quantum correlation and coherence are transient at high $g$ values. The maxima and minima of all measures are currently in good correspondence, resulting in strict agreement and validity. 
\subsection{Long range memory properties and the role of parameter $\alpha$}
\begin{figure}[b]
\includegraphics[width=16.5cm, height=14cm]{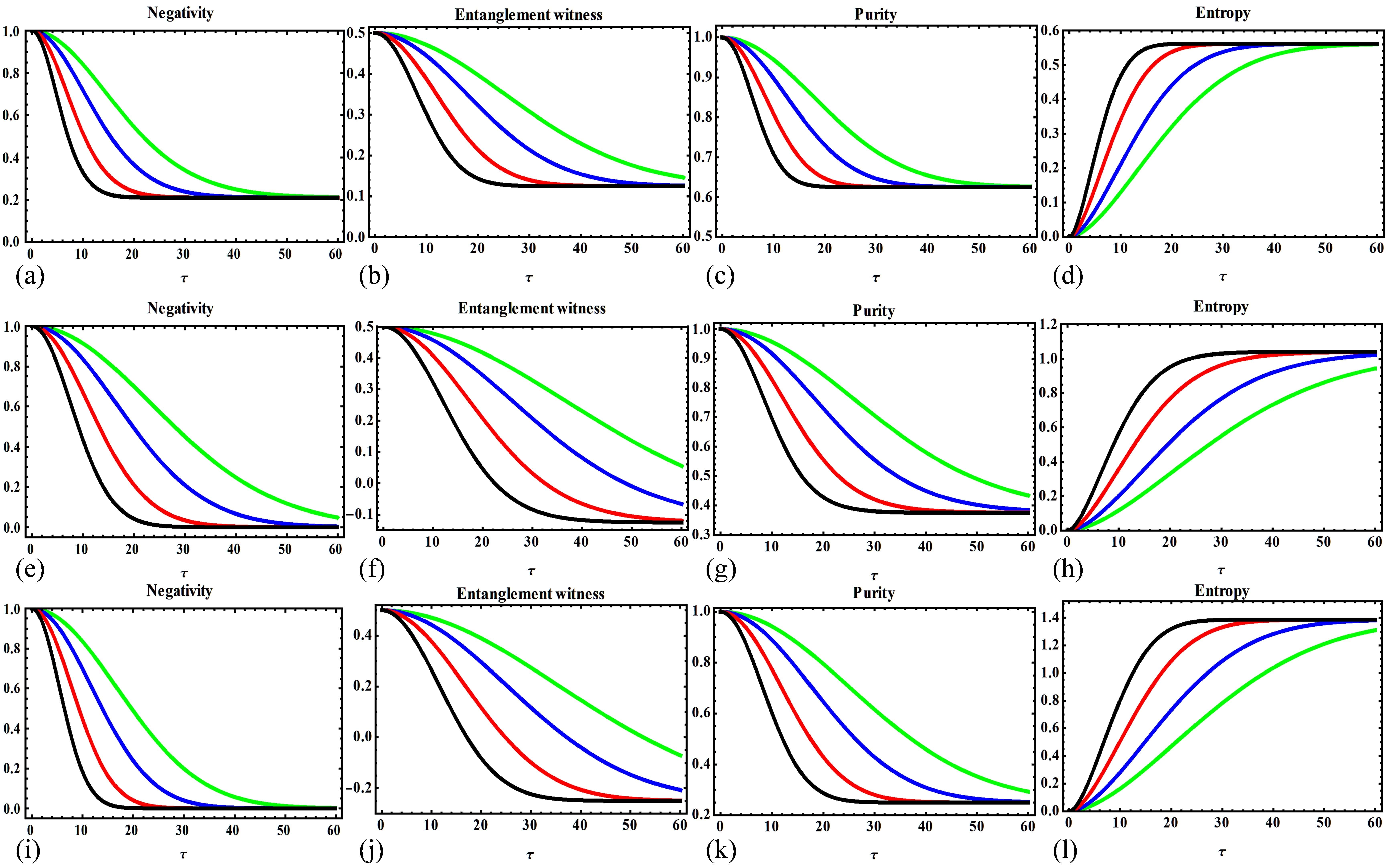}
\caption{Upper panel: Time evolution of negativty (a), entanglement witness (b), purity (c) and entropy (d) for the GHZ-like state when subjected to COM configuration under the effects of PL noise with the following parameter: $\alpha=3$ (green) $5$ (blue), $10$ (red), $20$ (black) with $g=10^{-4}$ and $p=1$. Middle panel: Same as upper panel but for the BIP configuration. Bottom panel: Same as upper panel but for the TRI configuration.}\label{Alpha figure}
\end{figure}
Fig.\ref{Alpha figure} depicts the role of the noisy parameter $\alpha$ in the dynamics of the GHZ-like state when coupled to COM (upper panel), BIP (middle panel), and TRI configurations under PL noise (bottom panel). In this case, $\alpha \neq 2$, which causes the $\beta$-function of the PL noise to become undefined. The current results show that $\alpha$ has a destructive effect on quantum correlation and coherence because the slopes shift from blue to black end as the parameter value increases, suggesting greater decay. The rest of the findings are close to the previous ones in Figs.\ref{negativity 3d graphs}, \ref{ewo 3d graphs}, \ref{purity 3d graphs} and \ref{decoherence 3d graphs}. It is now very clear that COM configuration remained a vital resource to preserve entanglement and coherence for an indefinite time under PL noise. In the BIP and TRI setups, quantum dynamics experienced faster and complete disentanglement and decoherence in a finite interval of time. We did not observe any revivals of the entanglement and coherence, even with different PL noise parameter settings. Hence, even at extremely low noise parameter values, PL noise significantly suppressed the ESD and ESB phenomena. Entanglement and coherence survival times were found to be longer than those obtained for bipartite and tripartite quantum systems subjected to Gaussian, PL noise, Ornstein Uhlenbeck, static, dynamics, and coloured noises specified in \cite{ATTA-GN, ATTA-PL, ATTA-PLFG, Kenfack, Zhang, Tchoffo, Buscemi, Lionel, Javed, Benedetti-color noise, Yu}. As a result, the existing manipulation of the noise parameter values stays an adequate setup for extended entanglement and coherence survival. This will significantly improve the quality of quantum computing protocols that require quantum information to be preserved over longer time intervals. Besides, we notice that the amount of decay is purely concerned with the number of PL noisy sources employed. According to $N(\tau)$, $EW(\tau)$ and $P(\tau)$, one can notice that the saturation levels are deeper in the cases where more than one noise source is deployed. Similarly, the decay levels of $EN(\tau)$ in a single noise source case are at shorter elevation than those having more than one PL noise sources. Following this, we conclude that the decay levels only depend upon the number of noisy sources applied and is not regulated by the parameter values. This contradicts our findings obtained for two qubit states where the decay levels are independent of the number of noisy sources subjected \cite{ATTA-PLFG}.
\section{Conclusion}\label{Conclusion}
Superconducting and solid state environments are widely used in practical reforms for effective functional quantum information processing. However, disorders in these circuits such as, thermal fluctuations and resistance led to failure and inefficiency of the quantum hardware and related protocols. These disorders primarily causes the non-local correlation and coherence to degrade or vanish completely. For preventing non-local correlations and coherence, detailed characterization of various noises and disorders, as well as the optimal configuration of noise parameters acting on quantum system subspaces, is critical. When subjected to classical fluctuating fields, we investigated the quantum correlation and coherence dynamics of three non-interacting qubits. The classical fields are believed to cause power-law (PL) noise, a type of Gaussian noise. This paper aimed to look into the negative effects of PL noise and classical environments in three different cases. For system environment couplings, we assumed common (COM), bipartite (BIP), and tripartite (TRI) configurations. We also distinguished between the dynamics of entanglement and coherence in the classical fields' PL noisy and noiseless realms. To clarify the degree of entanglement and coherence preserved by the system, various measures such as quantum negativity, entanglement witness, purity, and entropy were used. In addition, we present a detailed analysis of the purity factor of the GHZ-like state in the current situations. This will allow us to compare the initial quantum correlation and coherence encoded to that retained later in the system in order to understand the deterioration caused by PL noise.\\
Our findings show that the tripartite quantum correlation and coherence exhibit revival behaviour in isolated classical fields with no amplitude damping. When PL noise is superimposed on the joint phase of qubits and environment(s), entanglement and coherence degrade monotonically, exhibiting Markovian behaviour. Under Ornstein Uhlenbeck, fractional Gaussian, and pure Gaussian noises, the dynamics of bipartite and tripartite entangled states were found to be similarly decoherent \cite{Javed, Kenfack, ATTA-PLFG, ATTA-GN}, however, with varying preservation intervals and amounts. We realized that appropriate noise parameter fixation and system-environment coupling selection significantly reduced the noxious performance of the PL noise. To uphold non-local correlation and coherence over long intervals, the noise parameters $g$ and $\alpha$ of the PL noise should be kept as low as possible. The GHZ-like state successfully carried entanglement and coherence for indefinite or finite intervals in PL noisy assisted classical environments. It is worth noting that the GHZ-like state has higher preservation potential by remaining more stable in the presence of noise than the maximally entangled bipartite pure and mixture states, as well as the W-type of tripartite states investigated in \cite{Kenfack, Kenfack-static(p), Javed, ATTA-GN, ATTA-PLFG, Rossi}.\\
Due to PL noise in the COM configuration, the initial encoded quantum correlation and coherence in the GHZ-like state were not completely lost. Separability is achieved after a finite interaction time in all other cases eventually. No entanglement and coherence revivals were observed in the presence of PL noise, allowing quantum information to decay indefinitely and preventing information backflow between the system and the environment. This also confirms the inability of free states to transform into resource states in such PL noisy controlled environments. As a result, the initial encoded information in the GHZ-like states will be permanently lost and therefore cannot be recovered. This contradicts the dynamics of the entangled states in the thermal bath, as well as the dynamics of the qutrit-qutrit systems described in \cite{Zhang, Kenfack-static(p), Javed, Kenfack-GDE, Kenfack-qbqt, Zhang, Tchoffo, Buscemi, Lionel, Benedetti-color noise, Yu}.\\
The purity factor has been found to influence the retention power of GHZ-like states. Over a purity range defined as $0.85 \leq p \leq 1$, the GHZ-like state remained entangled and coherent. This is not the case for the purity range obtained using Ornstein Uhlenbeck and static noise for the same state, as shown in \cite{Kenfack, Kenfack-static(p)}. Besides that, in classical environments and PL noise, the maximum entanglement and coherence protection is only possible for states with $p=1$. The dynamical outlook character of the GHZ-like states at $p=1$ and other relative values is completely different.\\
Following that, common configuration proved to be the most competitive environment for increased quantum correlation and coherence protection. The maxima and minima of the measures are in good correspondence and have remained consistent in results demonstrating a strong connection between them. However, we discovered that quantum negativity, entanglement witness, and purity are more useful measures than entropy because they provide more precise information about the separability and decoherence of GHZ class states.\\
\section{Appendix}\label{Appendix}
This section represents the explicit density ensemble matrices for the dynamics of the tripartite GHZ-like state by using Eqs.\eqref{common environment}, \eqref{mixed environment} and \eqref{independent environment}. The statistical density mixture obtained for common system-environment coupling has the form as:
\begin{equation}
\rho^{CO}_{GHZ}=
\left[
\begin{array}{cccccccc}
 \nu_{CO}^{11} &  \nu_{CO}^{12} &  \nu_{CO}^{12} &  \nu_{CO}^{12} &  \nu_{CO}^{12} &  \nu_{CO}^{12} &  \nu_{CO}^{12} & \nu_{CO}^{11} \\
  \nu_{CO}^{12} &  \nu_{CO}^{13} &  \nu_{CO}^{13} &  \nu_{CO}^{13} &  \nu_{CO}^{13} &  \nu_{CO}^{13} &  \nu_{CO}^{13} &  \nu_{CO}^{12} \\
  \nu_{CO}^{12} &  \nu_{CO}^{13} &  \nu_{CO}^{13} &  \nu_{CO}^{13} &  \nu_{CO}^{13} &  \nu_{CO}^{13} &  \nu_{CO}^{13} &  \nu_{CO}^{12} \\
  \nu_{CO}^{12} &  \nu_{CO}^{13} &  \nu_{CO}^{13} &  \nu_{CO}^{13} &  \nu_{CO}^{13} &  \nu_{CO}^{13} &  \nu_{CO}^{13} &  \nu_{CO}^{12} \\
  \nu_{CO}^{12} &  \nu_{CO}^{13} &  \nu_{CO}^{13} &  \nu_{CO}^{13} &  \nu_{CO}^{13} &  \nu_{CO}^{13} &  \nu_{CO}^{13} &  \nu_{CO}^{12} \\
  \nu_{CO}^{12} &  \nu_{CO}^{13} &  \nu_{CO}^{13} &  \nu_{CO}^{13} &  \nu_{CO}^{13} &  \nu_{CO}^{13} &  \nu_{CO}^{13} &  \nu_{CO}^{12} \\
  \nu_{CO}^{12} &  \nu_{CO}^{13} &  \nu_{CO}^{13} &  \nu_{CO}^{13} &  \nu_{CO}^{13} &  \nu_{CO}^{13} &  \nu_{CO}^{13} &  \nu_{CO}^{12} \\
 \nu_{CO}^{11} &  \nu_{CO}^{12} &  \nu_{CO}^{12} &  \nu_{CO}^{12} &  \nu_{CO}^{12} &  \nu_{CO}^{12} &  \nu_{CO}^{12} & \nu_{CO}^{11}
\end{array}
\right]
\end{equation}
Where
\begin{align*}
\nu_{CO}^{11}=&\frac{5}{16}+\frac{3 \eta_{1}^{ew}}{16}, & \nu_{CO}^{12}=&-\frac{1}{16}+\frac{\eta_{1}^{ew}}{16},\\
\nu_{CO}^{13}=&\frac{1}{16}-\frac{\eta_{1}^{ew}}{16}.
\end{align*}
Next, for the mixed system-environment coupling, the explicit statistical density mixture reads as:
\begin{equation}
\rho^{GHZ}_{MX}=\left[
\begin{array}{cccccccc}
  \nu_{MX}^{14} & 0 &  \nu_{MX}^{15} & 0 & 0 &  \nu_{MX}^{15} & 0 &  \nu_{MX}^{14} \\
 0 &  \nu_{MX}^{16} & 0 &  \nu_{MX}^{16} &  \nu_{MX}^{16} & 0 &  \nu_{MX}^{16} & 0 \\
  \nu_{MX}^{15} & 0 &  \nu_{MX}^{17} & 0 & 0 &  \nu_{MX}^{17} & 0 &  \nu_{MX}^{15} \\
 0 &  \nu_{MX}^{16} & 0 &  \nu_{MX}^{16} &  \nu_{MX}^{16} & 0 &  \nu_{MX}^{16} & 0 \\
 0 &  \nu_{MX}^{16} & 0 &  \nu_{MX}^{16} &  \nu_{MX}^{16} & 0 &  \nu_{MX}^{16} & 0 \\
  \nu_{MX}^{15} & 0 &  \nu_{MX}^{17} & 0 & 0 &  \nu_{MX}^{17} & 0 &  \nu_{MX}^{15} \\
 0 &  \nu_{MX}^{16} & 0 &  \nu_{MX}^{16} &  \nu_{MX}^{16} & 0 &  \nu_{MX}^{16} & 0 \\
  \nu_{MX}^{14} & 0 &  \nu_{MX}^{15} & 0 & 0 &  \nu_{MX}^{15} & 0 &  \nu_{MX}^{14}
\end{array}
\right]
\end{equation}
Where
\begin{align*}
\nu_{MX}^{14}=&\frac{3}{16}+\frac{\eta_{1}^{ew}}{4}+\frac{\eta_{2}^{ew}}{16} ,&
\nu_{MX}^{15}=&-\frac{1}{16}+\frac{\eta_{2}^{ew}}{16},\\
\nu_{MX}^{16}=&\frac{1}{16}-\frac{\eta_{2}^{ew}}{16},&
\nu_{MX}^{17}=&\frac{3}{16}-\frac{\eta_{1}^{ew}}{4}+\frac{\eta_{2}^{ew}}{16}.
\end{align*}
Finally, the density matrix obtained for the independent system-environment coupling for the GHZ-like state is given as:
\begin{equation}
\rho^{GHZ}_{IN}=\left[
\begin{array}{cccccccc}
 \nu_{IN}^{18}  & 0 & 0 & 0 & 0 & 0 & 0 &  \nu_{IN}^{18} \\
 0 &  \nu_{IN}^{19} & 0 & 0 & 0 & 0 &  \nu_{IN}^{19} & 0 \\
 0 & 0 &  \nu_{IN}^{19} & 0 & 0 &  \nu_{IN}^{19} & 0 & 0 \\
 0 & 0 & 0 &  \nu_{IN}^{19} &  \nu_{IN}^{19} & 0 & 0 & 0 \\
 0 & 0 & 0 &  \nu_{IN}^{19} &  \nu_{IN}^{19} & 0 & 0 & 0 \\
 0 & 0 &  \nu_{IN}^{19} & 0 & 0 &  \nu_{IN}^{19} & 0 & 0 \\
 0 &  \nu_{IN}^{19} & 0 & 0 & 0 & 0 &  \nu_{IN}^{19} & 0 \\
  \nu_{IN}^{18} & 0 & 0 & 0 & 0 & 0 & 0 &  \nu_{IN}^{18}
\end{array}
\right]
\end{equation}
Where
\begin{align*}
\nu_{IN}^{18}=&\frac{1}{8}+\frac{3 \eta_{1}^{ew}}{8},& \nu_{IN}^{19}=&\frac{1}{8}-\frac{\eta_{1}^{ew}}{8}.
\end{align*}
\section{Acknowledgments}The present work is supported by the State Key Program of National Natural Science of China (Grant No. 61332019), the Major State Basic Research Development Program of China (973 Program, Grant No. 2014CB340601), the National Science Foundation of China (Grant No. 61202386 and No. 61402339).
\section{Data availability statement}
The data that support the findings of this study are available upon reasonable request from the
authors.

\end{document}